\documentclass[english,pra,showpacs,9pt]{revtex4-1}
\usepackage[T1]{fontenc}
\usepackage[latin9]{inputenc}
\usepackage{amsmath}
\usepackage{amssymb}
\usepackage{babel}

\begin{document}

\title{Comparison of mixed quantum states}

\author{Shengshi Pang and Shengjun Wu}

\affiliation{Hefei National Laboratory for Physical Sciences at Microscale and
Department of Modern Physics, University of Science and Technology
of China, Hefei, Anhui 230026, China}

\date{\today}
\begin{abstract}
In this article, we study the problem of comparing mixed quantum states:
given $n$ unknown mixed quantum states, can one determine whether
they are identical or not with an unambiguous quantum measurement?
We first study universal comparison of mixed quantum states, and prove
that this task is generally impossible to accomplish. Then, we focus
on unambiguous comparison of $n$ mixed quantum states arbitrarily
chosen from a set of $k$ mixed quantum states. The condition for
the existence of an unambiguous measurement operator which can produce
a conclusive result when the unknown states are actually the same
and the condition for the existence of an unambiguous measurement
operator when the unknown states are actually different are studied
independently. We derive a necessary and sufficient condition for
the existence of the first measurement operator, and a necessary condition
and two sufficient conditions for the second. Furthermore, we find
that the sufficiency of the necessary condition for the second measurement
operator has simple and interesting dependence on $n$ and $k$. At
the end, a unified condition is obtained for the simultaneous existence
of these two unambiguous measurement operators.
\end{abstract}


\maketitle

\section{Introduction}

Quantum identification problems such as discrimination of quantum
states \cite{chefles unambiguous 1,Dieks,duan lu ming,Ivanovic,Peres,shengshipang,discrimination mixed}
and discrimination of quantum operations (or quantum channels) \cite{discrimination operations 1,discrimination operations 2,discrimination operations 3,discrimination operations 4,discrimination operations 6}
are to determine the identity of an unknown quantum state (or quantum
channel) chosen from a given set. They have attracted broad interest
and received intensive research on the realization conditions and
the optimal implementation schemes, due to their key roles in some
important quantum information problems like quantum dense coding \cite{dense coding1,dense coding2}
and so on.

In recent years, a new problem relevant to quantum identification,
namely quantum state comparison, has become a new focus of interest
in the field of quantum information. Comparison of quantum states
is to find out whether several unknown quantum states are identical
or not. The reason that it receives much attention is that it has
been found especially useful in some interesting quantum verification
applications such as quantum fingerprinting \cite{fingerprinting,fingerprinting2}
and quantum digital signatures \cite{signature,signature2,signature3}.

The quantum state comparison problem seems similar with quantum state
identification at first sight, for if the unknown states can be identified
or discriminated, then one can immediately know whether they are identical
or different, but it is essentially different from the quantum state
identification problem: quantum state comparison only requires to
find whether the unknown states are the same or not but not the exact
identity of each unknown state, so it is not a must to employ a quantum
state identification protocol generally and, what's more, quantum
state comparison is usually a non-local operation and can be performed
on all the unknown states simultaneously.

So far, universal and unambiguous comparison of pure quantum states
has been studied in \cite{ensembles 08,jmo,orginal pla,unambiguous jpa,generalization not universal},
and optimization of the measurement strategy is considered in \cite{unambiguous jpa,generalization not universal}.
Comparison of quantum states has already been realized in experiments
\cite{coherent experimental}. Moreover, comparison of quantum operations
and quantum channels has also been investigated in recent research
\cite{comaprsion unitary channel,comparison measurements,comparison oracles}.

The previous researches on the quantum state comparison problem have
been largely focused on pure quantum states, and in this article,
we aim to study the comparison of mixed quantum states.

We shall study two types of the quantum state comparison problem in
this work. The first type is the universal comparison of mixed quantum
states, formulated as follows: given $n$ arbitrary unknown mixed
quantum states $\rho_{1},\cdots,\rho_{n}$, is it possible to determine
whether they are the same state or not? The second type is the unambiguous
comparison of mixed quantum states chosen from a finite set: given
$n$ unknown mixed quantum states $\rho_{1},\cdots,\rho_{n}$ arbitrarily
chosen from a finite fixed set consisting of $k$ mixed states $\sigma_{i}$
$(i=1,\cdots,k)$ (any state in the set is allowed to be chosen multiple
times), can one determine whether they are identical or not by an
unambiguous quantum measurement? {}``Unambiguous'' means that the
measurement may produce an inclusive result, but if the result is
conclusive it must be correct. For succinctness, the set of $k$ mixed
states $\{\sigma_{i}\}_{i=1}^{k}$ will be called \emph{candidate
set}, and the states $\sigma_{i}$ $(i=1,\cdots,k)$\emph{ }will be
called \emph{candidate states}. The study of unambiguous comparison
of mixed states will be based on a \emph{non-trivial condition} which
is presented in Sec. II.

In this article, we shall first study the universal comparison of
unknown mixed quantum states, and show that this task is impossible
to be achieved, in contrast to the universal comparison of pure quantum
states which is {}``half achievable'' \cite{unambiguous jpa} (the
meaning of {}``half achievable'' will be clear in Sec. III). Next,
we shall study the unambiguous comparison of unknown mixed quantum
states chosen from a given state set. This problem consists of two
parts: the possibility to obtain a conclusive result from an unambiguous
quantum measurement when the unknown states are actually the same
and the possibility to obtain a conclusive result from an unambiguous
quantum measurement when the unknown states are actually different.
It will be shown that the conditions to realize these two possibilities
are different. For the first possibility, we derive a necessary and
sufficient condition, and for the second, we obtain a necessary condition
and two sufficient conditions.

It is worth mentioning that in earlier researches on quantum comparison
of pure states, the main interest was in the condition of realizing
the first possibility. In this paper, we not only find a perfect necessary
and sufficient condition for the first possibility, but also get fruitful
results for the second. For instance, intuitively, the realizability
of quantum state comparison should have some dependence on the number
of states to be compared and the number of candidate states, but no
such result was obtained in any previous literature. However, in our
research on the second possibility, we derive such an interesting
result. Moreover, in previous researches, the conditions for the two
cases were separate and independent of each other, but in our article,
we shall establish a unified necessary and sufficient condition for
realizing both possibilities simultaneously.

\section{Preliminary}

Let us begin to present our work with the definitions of some notations
used in this paper. We assume that all states lie in a Hilbert space
$\mathcal{H}$ of finite dimension $d$ throughout this article. We
shall use the \emph{positive operator-valued measure} (POVM) formalism
to describe a comparison process. Generally, a POVM representation
of a physical process consists of a series of positive operators corresponding
to possible outcomes of the physical process, and these positive operators
must sum up to the identity operator. In our research, we shall use
$M_{1}$ to denote the POVM element which indicates that the $n$
states are identical, $M_{2}$ to denote the POVM element which indicates
that the $n$ states are different, and an additional operator $M_{?}$
to denote the POVM element which represents the inconclusive result.
Note that $M_{?}$ will only occur in the unambiguous comparison of
quantum states. We shall also use $M_{1}$, $M_{2}$ and $M_{?}$
to represent the corresponding measurement outcomes occasionally if
no unambiguity occurs in the contexts. The probability that $M_{i}$
occurs is
\begin{equation}
\mathrm{Prob}\left(M_{i}\right)=\mathrm{Tr}\left(M_{i}\rho_{1}\otimes\cdots\otimes\rho_{n}\right),\quad\forall i=1,2,?.\label{eq:0}
\end{equation}
 Throughout this article, we assume that the $n$ states are possessed
simultaneously so that collective measurements can be performed on
the whole $n$ states, and thus all the measurement operators $M_{i}$,
$i=1,\,2,\,?$, act on the composite Hilbert space $\mathcal{H}^{\otimes n}$.

In addition, we shall use $\mathrm{Supp}(\mathcal{O})$ to denote
the support of an operator $\mathcal{O}$, $\mathrm{Supp}^{\perp}(\mathcal{O})$
to denote the subspace orthogonal to $\mathrm{Supp}(\mathcal{O})$
in $\mathcal{H}$, and $\mathcal{O}$ can be a density operator or
a more general Hermitian operator. And ${\displaystyle \sum_{i}}\mathrm{Supp}(\mathcal{O}_{i})$
will be used to denote the subspace spanned by the supports of several
$\mathcal{O}_{i}$, i.e.
\[
\sum_{i}\mathrm{Supp}(\mathcal{O}_{i})=\Big\{\sum_{i}\overrightarrow{v_{i}}|\forall\overrightarrow{v_{i}}\in\mathrm{Supp}(\mathcal{O}_{i})\Big\}.
\]

It can be seen that the measurements like $M_{1}=0$ or $M_{2}=0$
can serve as trivial schemes for unambiguous comparison of mixed quantum
states. Obviously this kind of measurement should not be considered,
so it is appropriate to require that the conclusive result corresponding
to \emph{ }$M_{1}$ (or $M_{2}$) must occur with a non-zero probability
\emph{for at least one group of $n$ states}, and we call this condition
as \emph{non-triviality condition}. This condition will be held throughout
this article.

\section{Universal comparison of mixed states}

Our first result is devoted to universal comparison of mixed quantum
states. Universal comparison of quantum states concerns with states
arbitrarily chosen from the whole Hilbert space $\mathcal{H}$. In
the problem of universal comparison of pure quantum states, it is
known that it is impossible to produce an unambiguous result with
non-zero probability when the $n$ states are the same but possible
when the $n$ states are different \cite{orginal pla,jmo,unambiguous jpa},
i.e. $M_{1}$ does not exist but $M_{2}$ exists (and this is why
we said universal comparison of pure states was {}``half achievable''
in Sec. I). However, when the states to be compared are mixed, we
show below that it is also impossible to produce an unambiguous comparison
result when the states are different, i.e. $M_{2}$ does not exist
for universal comparison of mixed quantum states.

\emph{Theorem 1.} Universal comparison of unknown mixed quantum states
is impossible.

\emph{Proof.} We only need to prove that $M_{2}$ vanishes in the
case of mixed quantum states. Suppose $M_{2}$ exists for distinguishing
different mixed quantum states universally, then
\begin{equation}
\mathrm{Tr}\left(M_{2}\rho_{1}\otimes\rho_{2}\otimes\cdots\otimes\rho_{n}\right)=0\label{eq:1}
\end{equation}
 when $\rho_{1}=\rho_{2}=\cdots=\rho_{n}$ due to the unambiguity
of $M_{2}$.

Now we select arbitrary $n$ pure states $|\psi_{1}\rangle,\cdots,|\psi_{n}\rangle$
from the Hilbert space $\mathcal{H}$, and let
\begin{equation}
\rho_{1}=\rho_{2}=\cdots=\rho_{n}=\frac{1}{d}\left(|\psi_{1}\rangle\langle\psi_{1}|+\cdots+|\psi_{n}\rangle\langle\psi_{n}|\right).\label{eq:2}
\end{equation}
 Then
\begin{equation}
\mathrm{Prob}(R_{2})=\mathrm{Tr}\left(M_{2}\rho_{1}\otimes\rho_{2}\otimes\cdots\otimes\rho_{n}\right)=\frac{1}{d^{n}}\mathrm{Tr}\left(M_{2}\left(|\psi_{1}\rangle\langle\psi_{1}|+\cdots+|\psi_{n}\rangle\langle\psi_{n}|\right)^{\otimes n}\right)=0.\label{eq:3}
\end{equation}
 Eq. \eqref{eq:3} can be rewritten as
\begin{equation}
\sum_{i_{1,}\cdots,i_{n}=1}^{n}\mathrm{Tr}\left(M_{2}|\psi_{i_{1}}\rangle\langle\psi_{i_{1}}|\otimes\cdots\otimes|\psi_{i_{n}}\rangle\langle\psi_{i_{n}}|\right)=0.\label{eq:14}
\end{equation}
 Since $|\psi_{i_{1}}\rangle\langle\psi_{i_{1}}|\otimes\cdots\otimes|\psi_{i_{n}}\rangle\langle\psi_{i_{n}}|$
is an $n$-partite density operator, we have
\begin{equation}
\mathrm{Tr}\left(M_{2}|\psi_{i_{1}}\rangle\langle\psi_{i_{1}}|\otimes\cdots\otimes|\psi_{i_{n}}\rangle\langle\psi_{i_{n}}|\right)\geq0,\label{eq:4}
\end{equation}
 thus, with Eq. \eqref{eq:14}, there must be
\begin{equation}
\mathrm{Tr}\left(M_{2}|\psi_{i_{1}}\rangle\langle\psi_{i_{1}}|\otimes\cdots\otimes|\psi_{i_{n}}\rangle\langle\psi_{i_{n}}|\right)=0\label{eq:16}
\end{equation}
 for all $i_{1},i_{2},\cdots,i_{n}=1,\cdots,n$. Considering $|\psi_{1}\rangle,\cdots,|\psi_{n}\rangle$
are arbitrarily chosen from the Hilbert space $\mathcal{H}$, it can
be concluded that $M_{2}=0$. $\blacksquare$

Theorem 1 exhibits the difference between the universal comparison
of pure quantum states and that of mixed quantum states. For pure
quantum states, when they are the same the comparison measurement
can never give any conclusive result, and when they are different
the comparison measurement is possible to give a conclusive result
with non-zero probability. However, for mixed quantum states, the
universal comparison is always impossible.

\section{Unambiguous comparison of mixed states from a finite set}

In the previous section, we studied universal comparison of mixed
quantum states and showed that the task was impossible to accomplish.
In the following, we shall turn our attention to unambiguous comparison
of unknown mixed quantum states $\rho_{1},\cdots,\rho_{n}$ selected
from a finite set $\{|\sigma_{i}\rangle\}_{i=1}^{k}$. We divide this
problem into two parts and study them separately: the possibility
to produce an unambiguous result when the $n$ mixed states are actually
the same (i.e. the existence of $M_{1}$) and the possibility to produce
an unambiguous result when the $n$ mixed states are actually different
(i.e. the existence of $M_{2}$), as we shall see that the results
for these two possibilities are quite different.

We first study the first part, i.e. the existence of $M_{1}$. In
the following theorem, we show what condition the $k$ candidate states
$\sigma_{1},\cdots,\sigma_{k}$ should satisfy in order that an unambiguous
result can be produced when the $n$ states to be compared are identical.

\emph{Theorem 2.} For unambiguous comparison of mixed quantum states,
a conclusive result can be produced with a non-zero probability when
the $n$ states are actually the same, i.e., a non-trivial $M_{1}$
exists, if and only if
\begin{equation}
\mathrm{Supp}(\sigma_{i})\nsubseteq\sum_{j\neq i}\mathrm{Supp}\left(\sigma_{j}\right),\quad\exists i\in\{1,\cdots,k\}.\label{eq:28}
\end{equation}

\emph{Proof.} We first prove the {}``only if'' part by contradiction.
Assume that $\mathrm{Supp}(\sigma_{i})\subseteq{\displaystyle \sum_{j\neq i}\mathrm{Supp}\left(\sigma_{j}\right)}$
for all $i=1,\cdots,k$ and let $\rho_{1}=\cdots=\rho_{n}=\sigma_{i}$.
Since $\mathrm{Supp}(\sigma_{i})\subseteq{\displaystyle \sum_{j\neq i}}\mathrm{Supp}\left(\sigma_{j}\right)$,
we can always find positive coefficients $\alpha_{1},\cdots,\alpha_{i-1},\alpha_{i+1},\cdots,\alpha_{k}$
such that
\begin{equation}
\sum_{j\neq i}\alpha_{j}\sigma_{j}-\sigma_{i}\geq0,\label{eq:7}
\end{equation}
 where {}``$\geq$'' means semi-definite positive. Then we have
\begin{equation}
\mathrm{Tr}\Big(M_{1}\Big(\sum_{j\neq i}\alpha_{j}\sigma_{j}-\sigma_{i}\Big)\otimes\sigma_{i}\otimes\cdots\otimes\sigma_{i}\Big)\geq0,\label{eq:8}
\end{equation}
 or equivalently
\begin{equation}
\sum_{j\neq i}\mathrm{Tr}\Big(M_{1}\alpha_{j}\sigma_{j}\otimes\sigma_{i}\otimes\cdots\otimes\sigma_{i}\Big)\geq\mathrm{Tr}\left(M_{1}\sigma_{i}\otimes\cdots\otimes\sigma_{i}\right).\label{eq:9}
\end{equation}

According to the unambiguity of the measurement, we have
\begin{equation}
\mathrm{Tr}\left(M_{1}\sigma_{j}\otimes\sigma_{i}\otimes\cdots\otimes\sigma_{i}\right)=0,\quad\forall j\neq i.\label{eq:11}
\end{equation}
 Thus the left side of Eq. \eqref{eq:9} is equal to $0$ and
\begin{equation}
\mathrm{Tr}\left(M_{1}\sigma_{i}\otimes\cdots\otimes\sigma_{i}\right)\leq0,\;\forall i=1,\cdots,k.\label{eq:38}
\end{equation}
 On the other hand, $\mathrm{Tr}\left(M_{1}\sigma_{i}\otimes\cdots\otimes\sigma_{i}\right)\geq0$,
so it leads to
\begin{equation}
\mathrm{Tr}\left(M_{1}\sigma_{i}\otimes\cdots\otimes\sigma_{i}\right)=0,\;\forall i=1,\cdots,k.\label{eq:12}
\end{equation}

Note that when $\rho_{1},\cdots,\rho_{n}$ are arbitrary different
states from the set $\{|\sigma_{i}\rangle\}_{i=1}^{k}$, the unambiguity
of the measurement gives
\begin{equation}
\mathrm{Tr}\left(M_{1}\rho_{1}\otimes\cdots\otimes\rho_{n}\right)=0.\label{eq:39}
\end{equation}
 From \eqref{eq:12} and \eqref{eq:39}, it can be inferred that $M_{1}=0$.
Therefore the assumption $\mathrm{Supp}(\sigma_{i})\subseteq{\displaystyle \sum_{j\neq i}}\mathrm{Supp}\left(\sigma_{j}\right)$
for all $i=1,\cdots,k$ is false, and Eq. \eqref{eq:28} must be satisfied.

Now we prove the {}``if'' part of the theorem. Suppose $\mathrm{Supp}(\sigma_{i_{0}})\nsubseteq{\displaystyle \sum_{j\neq i_{0}}}\mathrm{Supp}\left(\sigma_{j}\right)$
for some $i_{0}\in\{1,\cdots,k\}$, it is easy to see that for each
state one can determine whether it is $\sigma_{i_{0}}$ or not by
local measurements: let $\mathrm{Proj}\Big({\displaystyle \sum_{j\neq i_{0}}}\mathrm{Supp}\left(\sigma_{j}\right)\Big)^{\perp}$
denote the projector onto the subspace orthogonal to ${\displaystyle \sum_{j\neq i_{0}}}\mathrm{Supp}\left(\sigma_{j}\right)$,
then
\[
\mathrm{Tr}\Big(\sigma_{l}\mathrm{Proj}\Big(\sum_{j\neq i_{0}}\mathrm{Supp}\left(\sigma_{j}\right)\Big)^{\perp}\Big)\neq0
\]
 only when $\sigma_{l}=\sigma_{i_{0}}$. So, let
\begin{equation}
M_{1}=\mathrm{Proj}^{\otimes n}\Big(\sum_{j\neq i_{0}}\mathrm{Supp}\left(\sigma_{j}\right)\Big)^{\perp},\label{eq:13}
\end{equation}
 then $M_{1}$ will produce a conclusive result when all the $n$
states are all $\sigma_{i_{0}}$. $\blacksquare$

It is worth mentioning that Proposition 1 in Ref. \cite{generalization not universal}
also gives the necessary and sufficient condition for the existence
of $M_{1}$ and it is similar to Theorem 2. Note that the difference
between {}``$\forall\, i$'' in Proposition 1 of \cite{generalization not universal}
and {}``$\exists i$'' in Theorem 2 results from that Proposition
1 in \cite{generalization not universal} requires all groups of unknown
states can be decided whether the states are identical.

\emph{Remark 1.} Readers who are familiar with the problem of quantum
state discrimination may find the conclusion of Theorem 2 is somewhat
similar to the condition of mixed quantum state discrimination \cite{discrimination mixed}.
Actually, $M_{1}$ was constructed as a product of local discrimination
of each state in the above proof of the sufficiency. However, this
does not mean that a quantum comparison process must be always implemented
by quantum state discrimination operations. In fact, a quantum comparison
process acts on the composite Hilbert space $\mathcal{H}^{\otimes n}$
and is usually non-local, so it cannot be generally decomposed into
several quantum state discrimination operations. Therefore, the necessity
of the condition \eqref{eq:28} cannot be straightforwardly derived
from the known conclusions for the discrimination of mixed quantum
state. In addition, Theorem 4 will explicitly show that on some occasions
unambiguous quantum state discrimination can never be accomplished
while unambiguous quantum state comparison can still succeed with
non-zero probability.

Next, we study under what condition a quantum measurement can give
the correct and conclusive result when the states are different, i.e.,
the existence of the measurement operator $M_{2}$. It will be interesting
to see that the existence of $M_{2}$ relies not only on the structure
of the $k$ states in the set, but also on $n$, the number of states
to be compared. We first give a necessary condition for the existence
of $M_{2}$ below.

\emph{Theorem 3.} For unambiguous comparison of mixed quantum states,
an unambiguous quantum measurement exists which can produce a conclusive
result with a non-zero probability when the $n$ states are actually
different, i.e., a non-trivial $M_{2}$ exists, only if
\begin{equation}
\sum_{j\neq i}\mathrm{Supp}\left(\sigma_{j}\right)\nsubseteq\mathrm{Supp}(\sigma_{i}),\quad\forall i=1,\cdots,k.\label{eq:22}
\end{equation}

\emph{Proof.} We prove the theorem by contradiction. Assume ${\displaystyle \sum_{j\neq i}}\mathrm{Supp}\left(\sigma_{j}\right)\subseteq\mathrm{Supp}(\sigma_{i})$
for some $i_{0}\in\{1,\cdots,k\}$, then
\begin{equation}
\mathrm{Supp}\left(\sigma_{j}\right)\subseteq\mathrm{Supp}(\sigma_{i_{0}}),\;\forall j=1,\cdots,k,\label{eq:40}
\end{equation}
 so
\begin{equation}
\mathrm{Supp}\left(\rho_{1}\otimes\cdots\otimes\rho_{n}\right)\subseteq\mathrm{Supp}(\sigma_{i_{0}}^{\otimes n}),\label{eq:30}
\end{equation}
 where $\rho_{1},\cdots,\rho_{n}$ are arbitrarily chosen from the
$k$ states in the candidate set. As $M_{2}$ is unambiguous, we have
\begin{equation}
\mathrm{Tr}\left(M_{2}\sigma_{i_{0}}^{\otimes n}\right)=0.\label{eq:31}
\end{equation}
 Since $M_{2}\geq0$, Eq. \eqref{eq:31} indicates that
\begin{equation}
\mathrm{Supp}(\sigma_{i_{0}}^{\otimes n})\perp\mathrm{Supp}\left(M_{2}\right),\label{eq:15}
\end{equation}
 then with Eq. \eqref{eq:30}, the probability that $M_{2}$ occurs
is
\begin{equation}
\mathrm{Prob}(M_{2})=\mathrm{Tr}\left(M_{2}\rho_{1}\otimes\cdots\otimes\rho_{n}\right)=0\label{eq:32}
\end{equation}
 for any $n$ different states $\rho_{i}$ ($i=1,\cdots,n$) chosen
from the candidate set. $\blacksquare$

\emph{Remark 2.} In Theorem 1, it has been shown that universal unambiguous
comparison of mixed quantum states is impossible. In fact, Theorem
1 can be linked to Theorem 3 and derived from it. The universal comparison
of mixed quantum states can be considered as a special case of unambiguous
comparison of mixed states from a particular candidate set, i.e.,
the complete set of all mixed states. To derive Theorem 1 from Theorem
3, one only needs to note that the support of the mixed state $\frac{1}{d}I$
($d$ is the dimension of $\mathcal{H}$) is the whole Hilbert space
$\mathcal{H}$ and it certainly contains the support of any other
mixed state, so the necessary condition in Theorem 3 is violated,
leading to the impossibility of universal unambiguous comparison of
mixed states.

It should be pointed out that Theorem 3 is a necessary condition but
not a sufficient condition for the existence of $M_{2}$. For example,
suppose $k=3$, and the three candidate states are
\begin{equation}
\sigma_{1}=\frac{1}{2}(|\psi_{1}\rangle\langle\psi_{1}|+|\psi_{2}\rangle\langle\psi_{2}|),\,\sigma_{2}=\frac{1}{2}(|\psi_{2}\rangle\langle\psi_{2}|+|\psi_{3}\rangle\langle\psi_{3}|),\,\sigma_{3}=\frac{1}{2}(|\psi_{1}\rangle\langle\psi_{1}|+|\psi_{3}\rangle\langle\psi_{3}|),\label{eq:26}
\end{equation}
 where $|\psi_{1}\rangle,\,|\psi_{2}\rangle,\,|\psi_{3}\rangle$ are
orthogonal to each other. Apparently, these three candidate states
satisfy the necessary condition \eqref{eq:22}, however, there is
no way to produce a conclusive unambiguous result with a non-zero
probability for any two different states chosen from \eqref{eq:26}:
the support of $M_{2}$ must be orthogonal to the supports of $\sigma_{1}\otimes\sigma_{1}$,
$\sigma_{2}\otimes\sigma_{2}$, $\sigma_{3}\otimes\sigma_{3}$ due
to the unambiguity of $M_{2}$, but the sum of the supports of $\sigma_{1}\otimes\sigma_{1}$,
$\sigma_{2}\otimes\sigma_{2}$, $\sigma_{3}\otimes\sigma_{3}$ is
the whole composite Hilbert space $\mathcal{H}\otimes\mathcal{H}$,
so $M_{2}=0$ when $n=2$. Thus, Eq. \eqref{eq:22} is not sufficient
for the existence of $M_{2}$.

However, it is interesting to note that $M_{2}$ does exist for the
above example when $n=3$. In contrast to the case $n=2,$ when $n=3$
we have
\begin{equation}
\mathrm{Supp}\left(\sigma_{1}^{\otimes3}\right)+\mathrm{Supp}\left(\sigma_{2}^{\otimes3}\right)+\mathrm{Supp}\left(\sigma_{3}^{\otimes3}\right)\neq\mathcal{H}^{\otimes3}\label{eq:36}
\end{equation}
 and in fact,
\begin{equation}
|\psi_{1}\rangle\otimes|\psi_{2}\rangle\otimes|\psi_{3}\rangle\perp\mathrm{Supp}\left(\sigma_{1}^{\otimes3}\right)+\mathrm{Supp}\left(\sigma_{2}^{\otimes3}\right)+\mathrm{Supp}\left(\sigma_{3}^{\otimes3}\right),\label{eq:37}
\end{equation}
 therefore, $M_{2}$ can be constructed as the projector onto the
subspace spanned by the state $|\psi_{1}\rangle\otimes|\psi_{2}\rangle\otimes|\psi_{3}\rangle$
and all permutations of its three factor states (e.g. $|\psi_{2}\rangle\otimes|\psi_{3}\rangle\otimes|\psi_{1}\rangle$
etc.), and this projector can unambiguously produce a conclusive result
with a non-zero probability when the three states from the candidate
set \eqref{eq:26} are different.

From the above example one can see that whether it is possible to
obtain an unambiguous result when the $n$ unknown mixed states are
different depends not only on the structures of the candidate states,
but also on the number of states to be compared, i.e., $n$. One may
wonder what value of $n$ can make $M_{2}$ exist, if the candidate
states already satisfy the necessary condition \eqref{eq:22}. The
following interesting theorem answers this question.

\emph{Theorem 4.} In the problem of unambiguous comparison of mixed
quantum states, if $n\geq k$, then \eqref{eq:22} is not only necessary
but also sufficient for the existence of a non-trivial $M_{2}$.

\emph{Proof.} The necessity has been shown in Theorem 3, and we only
need to prove the sufficiency here, provided $n\geq k$. The way to
prove the sufficiency of \eqref{eq:22} is to construct a non-trivial
$M_{2}$ that can produce a conclusive result with a non-zero probability
for at least one group of $n$ different states.

Generally speaking, the comparison process is a non-local quantum
measurement on the $n$ states, and it seems not so easy to construct
a qualified $M_{2}$. Anyway, we can assume that $M_{2}$ can be decomposed
into local measurements on the $n$ states and explore whether such
$M_{2}$ exists, and we finally find that such $M_{2}$ does exist.

Suppose
\begin{equation}
M_{2}=P_{1}\otimes\cdots\otimes P_{n},\label{eq:5}
\end{equation}
 where $P_{1},\cdots,P_{n}$ are projectors performed on the $n$
states respectively.

Let us remove any candidate state whose support is covered by the
support of another candidate state (if the supports of several states
are the same, then just keep one of them and remove the others), and
denote the remaining states as $\sigma_{1}^{\prime},\cdots,\sigma_{r}^{\prime}$.

Since the measurement is unambiguous, when all the $n$ states are
identical, $M_{2}$ should never occur, i.e.,
\begin{equation}
\mathrm{Prob}(M_{2})=\mathrm{Tr}\left(M_{2}\sigma_{i}^{\prime}\otimes\cdots\otimes\sigma_{i}^{\prime}\right)=\mathrm{Tr}\left(P_{1}\sigma_{i}^{\prime}\right)\cdots\mathrm{Tr}\left(P_{n}\sigma_{i}^{\prime}\right)=0,\;\forall i=1,\cdots,r.\label{eq:6}
\end{equation}
 Now for $i=1$, Eq. \eqref{eq:6} can hold if one $P_{j}$ satisfies
$\mathrm{Tr}\left(P_{j}\sigma_{1}^{\prime}\right)=0$, so one can
choose $j=1$ and then
\begin{equation}
P_{1}=\mathrm{Proj}\Big(\mathrm{Supp}^{\perp}(\sigma_{1}^{\prime})\Big),\label{eq:10}
\end{equation}
 where $\mathrm{Proj}\Big(\mathrm{Supp}^{\perp}(\sigma_{1})\Big)$
denotes the projector onto the subspace $\mathrm{Supp}^{\perp}(\sigma_{1})$.

Similarly, let
\begin{equation}
P_{2}=\mathrm{Proj}\Big(\mathrm{Supp}^{\perp}(\sigma_{2}^{\prime})\Big),\, P_{3}=\mathrm{Proj}\Big(\mathrm{Supp}^{\perp}(\sigma_{3}^{\prime})\Big),\cdots\label{eq:17}
\end{equation}
 then Eq. \eqref{eq:6} can hold for $i=2,\cdots,r$.

It is easy to see that if such construction can be repeated for all
$\sigma_{i}^{\prime}$ $(i=1,\cdots,r)$, the unambiguity condition
\eqref{eq:6} can be satisfied by all $\sigma_{i}^{\prime}$, and
also by all the removed states because the support of any removed
state is covered by one $\sigma_{i}^{\prime}$ $(i=1,\cdots,r)$.
Thanks to $n\geq k$, it is indeed possible to make the $n$ projectors
$P_{1},\cdots,P_{n}$ run over
\begin{equation}
\mathrm{Proj}\Big(\mathrm{Supp}^{\perp}(\sigma_{i}^{\prime})\Big),\; i=1,\cdots,r,\label{eq:18}
\end{equation}
 so $M_{2}$ can be constructed as
\begin{equation}
M_{2}=\bigotimes_{i=1}^{r}\mathrm{Proj}\Big(\mathrm{Supp}^{\perp}(\sigma_{i}^{\prime})\Big)\bigotimes I^{\otimes\left(n-r\right)}.\label{eq:27}
\end{equation}

Eq. \eqref{eq:27} can ensure that $M_{2}$ is not trivial, because
according to \eqref{eq:22}, for each $\sigma_{i}$ there is at least
one $\sigma_{j_{i}}$ such that
\begin{equation}
\mathrm{Supp}(\sigma_{j_{i}})\nsubseteq\mathrm{Supp}(\sigma_{i}),\label{eq:20}
\end{equation}
 and when the $n$ states containing $\sigma_{j_{1}},\cdots,\sigma_{j_{k}}$,
$\mathrm{Prob}(M_{2})>0$. $\blacksquare$

Theorem 4 is quite interesting, because in a general quantum state
comparison problem the number of the states chosen from a set can
be varied and it is natural to think that the number of the chosen
states may affect the possibility of the unambiguous comparison task.
However, in previous literatures, no such conditions were found, and
Theorem 4 is the first result to reveal the relation between the number
of the states to be compared and the possibility of the unambiguous
comparison task.

\emph{Remark 3.} Theorem 4 shows the difference between the quantum
state comparison problem and the quantum state discrimination problem.
Eq. \eqref{eq:22} is not a sufficient condition for unambiguous quantum
state discrimination, so the states which satisfies \eqref{eq:22}
are necessary to be unambiguously distinguishable, but Theorem 4 tells
that only if $n\geq k$, the states chosen from them can still be
unambiguously compared when they are actually different. And Eq. \eqref{eq:26}
is such an example: apparently the states in Eq. \eqref{eq:26} cannot
be distinguished in an unambiguous way, but when $n=3$, $M_{2}$
does exist indeed and it can be chosen as ${\displaystyle \sum_{\mathrm{Per}\in S}\mathrm{Per}(|\psi_{1}\rangle\langle\psi_{1}|\otimes|\psi_{2}\rangle\langle\psi_{2}|\otimes|\psi_{3}\rangle\langle\psi_{3}|)}$.

Next, we give another sufficient condition for the existence of $M_{2}$
which depends only on the structures of the candidate states, and
this sufficient condition holds true for all $n\geq2$.

\emph{Theorem 5.} For unambiguous comparison of mixed quantum states,
if the $k$ candidate states $\sigma_{1},\cdots,\sigma_{k}$ satisfy
\eqref{eq:22} and the following condition
\begin{equation}
\mathrm{Supp}(\sigma_{i_{0}})\nsubseteq\sum_{j\neq i_{0}}\mathrm{Supp}\left(\sigma_{j}\right),\quad\exists i_{0}\in\{1,\cdots,k\},\label{eq:23}
\end{equation}
 then a non-trivial $M_{2}$ exists.

\emph{Proof.} Like the proof of the last theorem, we construct a proper
$M_{2}$ starting from a product of local projectors on the $n$ states
to be compared to prove this theorem.

Suppose
\begin{equation}
M_{2}=P_{1}\otimes\cdots\otimes P_{n},\label{eq:21}
\end{equation}
 where $P_{1},\cdots,P_{n}$ are projectors on the $n$ states respectively.
Since the measurement is unambiguous, when $\rho_{1}=\cdots=\rho_{n}=\sigma_{i_{0}}$,
$M_{2}$ should never produce a conclusive result, implying
\begin{equation}
\mathrm{Prob}(M_{2})=\mathrm{Tr}\left(M_{2}\sigma_{i_{0}}\otimes\cdots\otimes\sigma_{i_{0}}\right)=\mathrm{Tr}\left(P_{1}\sigma_{i_{0}}\right)\cdots\mathrm{Tr}\left(P_{n}\sigma_{i_{0}}\right)=0,\label{eq:25}
\end{equation}
 so one can let $P_{1}=\mathrm{Proj}\biggl(\mathrm{Supp}^{\perp}(\sigma_{i_{0}})\biggr)$
and thus Eq. \eqref{eq:25} is met for $\sigma_{i_{0}}$. But such
a method cannot be repeated for other $j\neq\sigma_{i_{0}}$, because
$n$ may be smaller $k$ and in this case $P_{1},\cdots,P_{n}$ cannot
cover all $k$ projectors $\mathrm{Proj}\left(\mathrm{Supp}^{\perp}(\sigma_{j})\right),\; j=1,\cdots,k.$

Nevertheless, we can use another method to construct proper $P_{2},\cdots,P_{n}$
to ensure $\mathrm{Prob}(M_{2})=0$ when $\rho_{1}=\cdots=\rho_{n}=\sigma_{j},\, j\neq i_{0}$.
Considering Eq. \eqref{eq:23}, we have
\begin{equation}
\mathrm{Proj}\Big(\sum_{j\neq i_{0}}\mathrm{Supp}\left(\sigma_{j}\right)\Big)^{\perp}\neq0.\label{eq:33}
\end{equation}
 Note that
\[
\mathrm{Supp}\left(\sigma_{l}\right)\perp\Big(\sum_{j\neq i_{0}}\mathrm{Supp}\left(\sigma_{j}\right)\Big)^{\perp},\;\forall l\neq i_{0},
\]
 so
\begin{equation}
\mathrm{Tr}\Big(\sigma_{l}\mathrm{Proj}\Big(\sum_{j\neq i_{0}}\mathrm{Supp}\left(\sigma_{j}\right)\Big)^{\perp}\Big)=0,\;\forall l\neq i_{0}.\label{eq:34}
\end{equation}
 Then if one of $P_{2},\cdots,P_{n}$, say $P_{2}$, is $\mathrm{Proj}\Big({\displaystyle \sum_{j\neq i_{0}}}\mathrm{Supp}\left(\sigma_{j}\right)\Big)^{\perp}$,
the unambiguity requirement can be satisfied for all $\rho_{1}=\cdots=\rho_{n}=\sigma_{j},\, j\neq i_{0}$.
Thus, a proper $M_{2}$ can be constructed as follows,
\begin{equation}
M_{2}=\mathrm{Proj}\Big(\mathrm{Supp}^{\perp}(\sigma_{i_{0}})\Big)\otimes\mathrm{Proj}\Big(\sum_{j\neq i_{0}}\mathrm{Supp}\left(\sigma_{j}\right)\Big)^{\perp}\otimes I^{\otimes\left(n-2\right)}.\label{eq:24}
\end{equation}

And $\mathrm{Prob}(M_{2})\neq0$ when $\rho_{1},\cdots,\rho_{n}$
are different and contain $\sigma_{i}$, thus the non-triviality condition
is also satisfied. $\blacksquare$

From Theorem 2 to Theorem 5, we have studied the possibility to unambiguously
compare the $n$ unknown states when the $n$ states are the same
or different separately. However, it is interesting to note that the
sufficient condition \eqref{eq:23} for $M_{2}$ is essentially the
same as the necessary and sufficient condition \eqref{eq:28} for
$M_{1}$, so we can combine Theorems 2, 3, 5, and obtain the following
corollary on the possibility of an unambiguous quantum measurement
which can deal with both cases that the states are the same and that
the states are different.

\emph{Corollary 1. }For the problem of unambiguous comparison of mixed
quantum states from a given state set, an unambiguous measurement
exists which can produce conclusive results in both cases that the
$n$ states are actually the same or different with a non-zero probability
(i.e., non-trivial $M_{1}$ and $M_{2}$ exist simultaneously), if
and only if the following two conditions are both satisfied

\begin{equation}
\begin{aligned}\mathrm{i}. & \quad\sum_{j\neq i}\mathrm{Supp}\left(\sigma_{j}\right)\nsubseteq\mathrm{Supp}(\sigma_{i}),\;\forall i=1,\cdots,k;\\
\mathrm{ii.} & \quad\mathrm{Supp}(\sigma_{i})\nsubseteq\sum_{j\neq i}\mathrm{Supp}\left(\sigma_{j}\right),\;\exists i\in\{1,\cdots,k\}.
\end{aligned}
\label{eq:35}
\end{equation}

It is worth pointing out that most previous researches on the quantum
state comparison problem dealt with the two cases that the $n$ states
are the same or different separately, and most conclusions were also
separate for these two cases. Corollary 1 is the first unified necessary
and sufficient condition to deal with both two cases simultaneously.

\section{Conclusion}

In summary, we have studied the problem of comparing mixed quantum
states in this article. Two types of state comparison are covered:
universal comparison of mixed quantum states, and unambiguous comparison
of mixed states chosen from a given state set. The universal comparison
of mixed quantum states is shown to be generally impossible, in marked
contrast to the universal comparison of pure quantum states. The problem
of comparing mixed states from a given set is divided into two parts:
the condition for obtaining an unambiguous result when the states
to be compared are actually the same and the condition when the states
to be compared are actually different. We obtain a necessary and sufficient
condition for the first case, and a necessary condition and two sufficient
conditions for the second. An interesting theorem is derived for the
necessary condition of the second case to be sufficient dependent
on the relation between the number of the states to be compared and
the number of the states in the candidate set. Furthermore, we find
a unified necessary and sufficient condition for the the existence
of an unambiguous quantum measurement which can be effective in both
cases.

Comparison of mixed quantum states is an interesting and important
problem since it can be used in various quantum verification applications
like the quantum digital signature protocol, quantum fingerprinting,
and so on. We hope our work could contribute to a deeper understanding
and further solution to this problem.

\section*{Acknowledgement}

This work is supported by the NNSF of China (Grant No. 11075148),
the CUSF, the CAS and the National Fundamental Research Program. S.
Pang also acknowledges the support from the Innovation Foundation
of USTC.

\end{document}